\documentclass[%
reprint,
superscriptaddress,
 amsmath,amssymb,
 aps,
longbibliography,
floatfix,
]{revtex4-2}

\usepackage[dvipsnames,svgnames,x11names,hyperref]{xcolor}
\usepackage{siunitx}
\usepackage{tabularx}
\usepackage{color}
\usepackage{graphicx}
\usepackage{dcolumn}
\usepackage{bm}
\usepackage{hyperref}
\usepackage[mathlines]{lineno}
\usepackage{mathrsfs}
\usepackage{braket}
\usepackage{placeins} 
\usepackage{float} 
\usepackage{color}

\newcommand{\hide}[1]{}

\hypersetup{colorlinks=true, 
	linkcolor={blue!75!black!80!yellow},
	citecolor={blue!75!black!80!yellow}, 
	urlcolor=black 
	}
\makeatletter \renewcommand\@make@capt@title[2]{%
\@ifx@empty\float@link{\@firstofone}{\expandafter\href\expandafter{\float@link}}%
\sffamily{\textbf{#1}}\@caption@fignum@sep#2 }

\usepackage[normalem]{ulem}
\usepackage{color}

\begin{document}

\title{Chemical reactivity under collective vibrational strong coupling}

\author{Derek S. Wang}
\altaffiliation{Contributed equally to this work}
\affiliation{Harvard John A. Paulson School of Engineering and Applied Sciences, Harvard University, Cambridge, MA 02138, USA}

\author{Johannes Flick}
\altaffiliation{Contributed equally to this work}
\email{jflick@flatironinstitute.org}
\affiliation{Center for Computational Quantum Physics, Flatiron Institute, New York, NY 10010, USA}

\author{Susanne F. Yelin}
\email{syelin@g.harvard.edu}
\affiliation{Department of Physics, Harvard University, Cambridge, MA 02138, USA}

\begin{abstract}
\noindent Recent experiments of chemical reactions in optical cavities have shown great promise to alter and steer chemical reactions but still remain poorly understood theoretically. In particular the origin of resonant effects between the cavity and certain vibrational modes in the collective limit is still subject to active research. In this paper, we study unimolecular dissociation reactions of many molecules collectively interacting with an infrared cavity mode through their vibrational dipole moment.  We find that the reaction rate can slow down by increasing the number of aligned molecules if the cavity mode is resonant with a vibrational frequency of the molecules. We also discover a simple scaling relation that scales with the collective Rabi splitting to estimate the onset of reaction rate modification by collective vibrational strong coupling and numerically demonstrate these effects for up to $10^4$ molecules. 
\end{abstract}
\date{\today}

\maketitle

\section{Introduction} \label{sec:intro}
Recent studies in the field of vibrational polariton chemistry suggest that chemical reactivity---such as reaction rates \cite{Thomas2016}, product selectivity \cite{Thomas2019}, and charge-transfer equilibrium \cite{Pang2020}---can be modified by coupling the dipole moments of molecular vibrational modes to the vacuum electric field of infrared cavity modes. Despite intense efforts to develop a theory of vibrational polariton chemistry (see \textit{e.g.} Refs.~\citenum{Wang2021Perspective, Galego2019, Campos-Gonzalez-Angulo2019, Li2020a, Li2021a, Li2021, Schafer2021, Du2021, Yang2021, Mandal2021, Wang2021CavityIVR, Ahn2022, Lindoy2022, Sun2022, Philbin2022, Vendrell2022}), a comprehensive explanation remains elusive.

While assuming uncoupled harmonic molecular vibrational modes is computationally convenient for analytical and numerical models, chemical reactions, such as dissociation, inherently involve full exploration of anharmonic and coupled vibrational potentials that can profoundly impact reaction dynamics \cite{Hernandez2019, Triana2020}. In fact, recent computational studies \cite{Schafer2021, Wang2021CavityIVR} using \textit{ab initio} or classical models of chemical reactivity in a cavity show that cavity coupling can modify reaction dynamics by interfering with intramolecular vibrational energy redistribution (IVR), feasible only when the coupled and anharmonic nature of vibrational modes is taken into account.

In these studies, however, only a single molecule coupled to a cavity has been considered. To match the single-molecule Rabi splitting with the collective Rabi splitting observed in experiment, these theoretical studies require vacuum electric fields, leading to the light-matter coupling strength being much larger than experimentally realizable in optical cavities. It is still an open question how insights from predictions for single-molecule scenarios translate to the experimentally relevant case of many molecules. On the other hand, the development of cavity molecular dynamics simulations
\cite{Li2020b} 
that consider up to thousands of molecules has yielded novel insights into cavity-modified excitation and relaxation dynamics but not yet addressed changes in chemical reactivity.

In this work, we study changes in the chemical reactivity when many molecules are collectively coupled to an infrared cavity. We find that the reaction rate can be slowed down when the optical cavity mode is tuned in resonance with the vibrational mode. In this case, the collective vibrational state of the many molecules drives the cavity displacement, enabling cavity-mediated vibrational energy distribution as fast as IVR processes. We find that these effects may be observable in current experimental setups when the molecules are aligned and may be validated with measurements of molecule density-dependent cavity resonance frequencies. We also note that when we consider randomly oriented molecules, changes to the chemical reactivity largely disappear due to light-matter coupling that scales more weakly with the number of molecules than in the aligned case, suggesting further investigations are necessary to better understand the origin of cavity-modified ground state chemical reactivity.

\section{Model of many molecules in a cavity}

\begin{figure}[!tbhp]
\centering
\includegraphics[width=\linewidth]{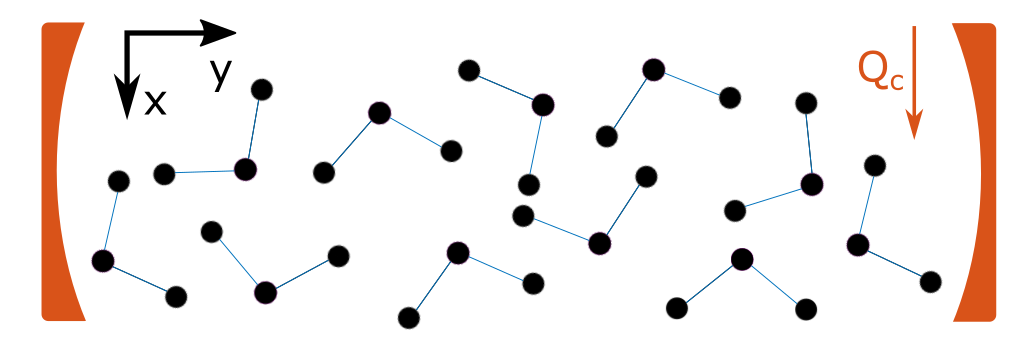}
\caption{Schematic of $N_\mathrm{mol}$ molecules coupled to an infrared cavity. Each molecule $n$ comprises three local vibrational modes: anharmonic stretch $q_{n,1}$, anharmonic stretch $q_{n,2}$, and harmonic bend $q_{n,3}$. We fix the orientation $\theta_n$ between the cavity displacement $Q_\mathrm{c}$ and the molecular axis.}
\label{fig:schematic}
\end{figure}

In this work, we study a simple, classical model for the vibrational dynamics of $N_\mathrm{mol}$ molecules collectively coupled to an optical cavity, taking into account the anharmonicity of the chemical bonds. We schematically illustrate this model in Fig. \ref{fig:schematic}. The total Hamiltonian can be written as $H = H_\mathrm{mol} + H_\mathrm{F}$, where the molecular Hamiltonian $H_\mathrm{mol}$ and field Hamiltonian $H_\mathrm{F}$ are  given by
\begin{subequations}
\begin{align}
    H_\mathrm{mol} &= \sum_{n=1}^{N_\mathrm{mol}} \Bigg[\sum_{i=1}^3\left(\frac{1}{2}G_{ii}^{(0)}p_{n,i}^2 + V_{n,i}(q_{n,i})\right) \\ \nonumber
    &+ \sum_{i,j}^3  G_{ij}^{(0)} p_{n,i} p_{n,j}, \Bigg], \\
    H_\mathrm{F} &= \sum_{k} \left[\frac{1}{2}P_{k}^2 +\frac{1}{2}\omega_k^2 \left(Q_{k}-\frac{\lambda_k}{\omega_k}\bm{\mu} \cdot \bm{\xi}\right)^2\right].
\end{align}
\end{subequations}

In the molecular Hamiltonian $H_\mathrm{mol}$, each molecule $n$ is represented by a classical triatomic model with three degrees of freedom ($q_{n,i}$) and anharmonic potentials ($V_{n,i}$). Previously used to describe the emergence of anharmonicity and chaos in IVR \cite{Bunker1962, Bunker1964,Karmakar2020}, this model was extended recently to describe chemical reactivity of a single molecule in an infrared optical cavity \cite{Wang2021CavityIVR}. We briefly review the molecular model: $i = \{1, 2\}$ correspond to the local stretching modes, and $i=3$ is the local bending mode. For the two stretching modes, the vibrational potentials are approximated by Morse potentials ($V_{n,i} = D\left\lbrace 1-\exp{\left[-\alpha_i \left(q_{n,i} - q_i^0\right)\right]}\right\rbrace^2$), where $D$ is the dissociation energy and $q_i^0$ the equilibrium position. For the bending mode, the vibrational potential is assumed to be harmonic: $V_{n,3} = \omega_3^2 (q_{n,3} - q_3^0)^2/(2G_{33}^{(0)})$. The definitions and numerical values for these parameters are provided in Refs. \citenum{Karmakar2020, Wang2021CavityIVR}. Notably, the local vibrational modes $q_{n,1}, q_{n,2}$, and $q_{n,3}$ are coupled through the vibrational momentum mode-mode coupling terms $G_{ij}^{(0)}$, resulting in three normal modes (normal bending, symmetric stretching, and anti-symmetric stretching).

In the field Hamiltonian $H_\mathrm{F}$, the variables $\omega_\mathrm{c}$, $Q_\mathrm{c}$, $P_\mathrm{c}$, $\lambda_\mathrm{c}$ are the frequency, displacement, momentum, and strength of the cavity mode ``c"; $\bm{\mu}=\sum_{n=1}^{N_\mathrm{mol}} \bm{\mu}_n=\mu_x\hat{x}+\mu_y\hat{y}$ is the permanent dipole moment of the $N_\mathrm{mol}$ molecules; $\bm{\xi}$ is the unit vector of polarization of the electric field; and $\theta_n$ is the angle between the molecular axis and $x$-axis. The parametrization of the permanent dipole moment $\bm{\mu}_n$ is described in Appendix \ref{app:dipolemoment}. 

Finally, to focus on the direct molecule-cavity coupling, we neglect potentially relevant molecular translations and rotations, intermolecular dipole-dipole and dispersion interactions, and photon loss through the cavity, respectively, although the presence of the cavity may impact these other terms as well.

We can then stably propagate the equations of motion derived from $H$ with Hamilton's equations using the 8th order Runge-Kutta method \cite{Dormand1981}. Because the dissociation dynamics depend on the initial state, we average over many ensembles, generated as described in Appendix \ref{app:initialstates}, for the results presented here. Briefly, in the main text, each molecule in an ensemble of $N_\mathrm{mol}$ molecules is initialized in a randomly chosen state with a given energy. Other initializations are possible; for instance, we explore the effect of coherently initializing all molecules in an ensemble with the same initial state in the Appendix. With this approach, in Ref.~\citenum{Wang2021CavityIVR} for the single-molecule case, we find that when the optical cavity is resonantly coupled to particular anharmonic vibrational modes, the cavity can interfere with IVR and alter unimolecular  dissociation reaction rates by acting as a reservoir for vibrational energy.

\section{Dissociation under collective coupling}

\begin{figure}[!tbhp]
\centering
\includegraphics[width=\linewidth]{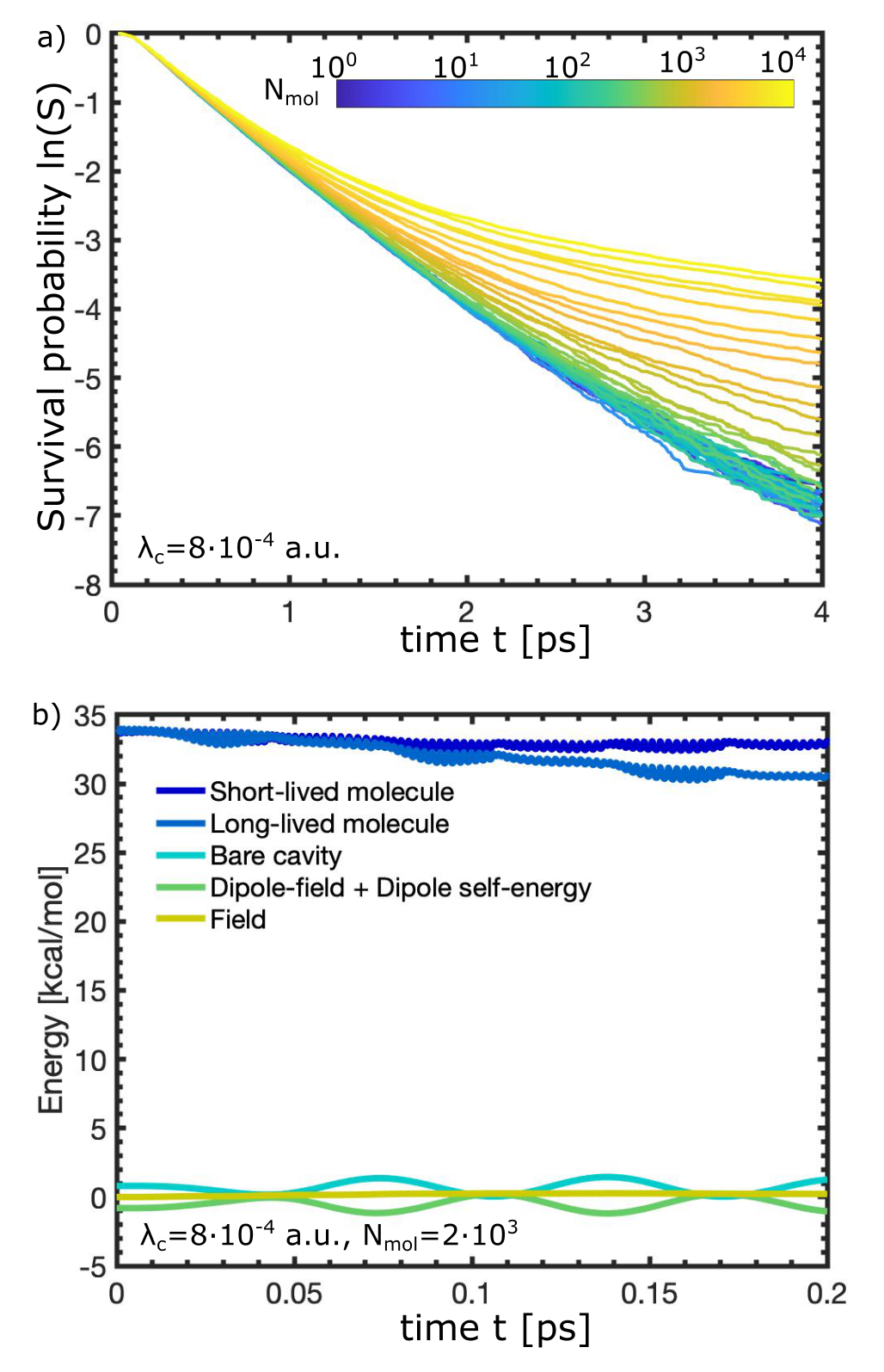}
\caption{\textbf{(a)} Reaction progress measured with time-dependent survival probability $S$. Increasing the number of molecules $N_\mathrm{mol}$ for constant cavity strength $\lambda_\mathrm{c}$ and cavity frequency $\omega_\mathrm{c}=520$ cm$^{-1}$ decreases the reaction rate for $t > 0.5$ ps. \textbf{(b)} Energy distribution within the first 0.2 ps for $\lambda_\mathrm{c}=8 \cdot 10^{-4}$ a.u. for $N_\mathrm{mol}=2\cdot 10^3$.
The change in the reaction rate can be attributed to cavity-mediated vibrational energy transfer, where decreases in molecular energy result in longer dissociation times or slower reaction rates, and vice versa.}
\label{fig:Nmol}
\end{figure}

We now investigate how chemical reactivity changes under collective vibrational strong coupling. We first initialize ensembles of $N_\mathrm{mol}$ molecules, where each molecule has an initial energy of 34 kcal/mol $> D$, where $D = 24$ kcal/mol is the dissociation energy, and then we plot the time-dependent survival probability $S$, or the ratio between the number of undissociated molecules and total molecules, in Fig. \ref{fig:Nmol}(a) for $N_\mathrm{mol}$ up to $10^4$. Importantly, we assume that all molecules are aligned along the cavity field on the $x$-axis with $\lambda_\mathrm{c}=8\cdot 10^{-4}$ a.u. and cavity frequency $\omega_\mathrm{c}=520$ cm$^{-1}$. The chosen cavity frequency $\omega_\mathrm{c}$ results in the maximum change in chemical reactivity for this molecular model when $N_\mathrm{mol}=1$ and $\theta=0$ \cite{Wang2021CavityIVR}. We find that the reaction slows down drastically for $N_\mathrm{mol}> 10^2-10^3$, and that this slowdown becomes more pronounced with increasing $N_\mathrm{mol}$.

To understand how the cavity changes the reaction rate under collective vibrational strong coupling, in Fig. \ref{fig:Nmol}(b) for the same $\lambda_\mathrm{c} = 8 \cdot 10^{-4}$ a.u., $\omega_\mathrm{c}=520$ cm$^{-1}$, and $N_\mathrm{mol}=2 \cdot 10^3$ we plot the distribution of energy per molecule of the cavity-molecule system within the first 0.2 ps for the following degrees of freedom: the average energy per short-lived ($0.5 < t_\mathrm{dis} < 1$ ps) or long-lived ($t_\mathrm{dis}>4$ ps) molecule, the bare cavity energy $(\frac{1}{2}P_\mathrm{c}^2+\frac{1}{2}\omega_\mathrm{c}^2 Q_\mathrm{c}^2)/N_\mathrm{mol}$, the dipole-field and dipole-field energy $(-\omega_\mathrm{c} \lambda_\mathrm{c} \mu_x Q_\mathrm{c}+\lambda_\mathrm{c}^2\mu_x^2/2)/N_\mathrm{mol}$, and the total field energy $H_\mathrm{F}/N_\mathrm{mol}$. The average energy of short-lived molecules decreases only slightly, while the average energy of long-lived molecules drops a couple of kcal/mol. Oscillations in the bare cavity energy are correlated with oscillations in the average energies of both short- and long-lived molecules, demonstrating femtosecond-scale energy exchange between the cavity and molecules. Just as in the single-molecule case \cite{Wang2021CavityIVR}, the cavity serves as a reservoir of vibrational energy. This process can be mechanistically explained as follows: as the $N_\mathrm{mol}$ molecules vibrate, they displace the cavity displacement $Q_\mathrm{c}$, the magnitude of which increases with $N_\mathrm{mol}$ for a given $\lambda_\mathrm{c}$. Larger $Q_\mathrm{c}$ magnitudes enable transiently larger forces on the molecular coordinates that interfere with the bare intramolecular vibrational dynamics. Overall, here we explicitly observe that vibrational strong coupling between many molecules and a cavity mode can augment energy transfer between molecules through the collectively driven cavity displacement.

\begin{figure}[!tbhp]
\centering
\includegraphics[width=\linewidth]{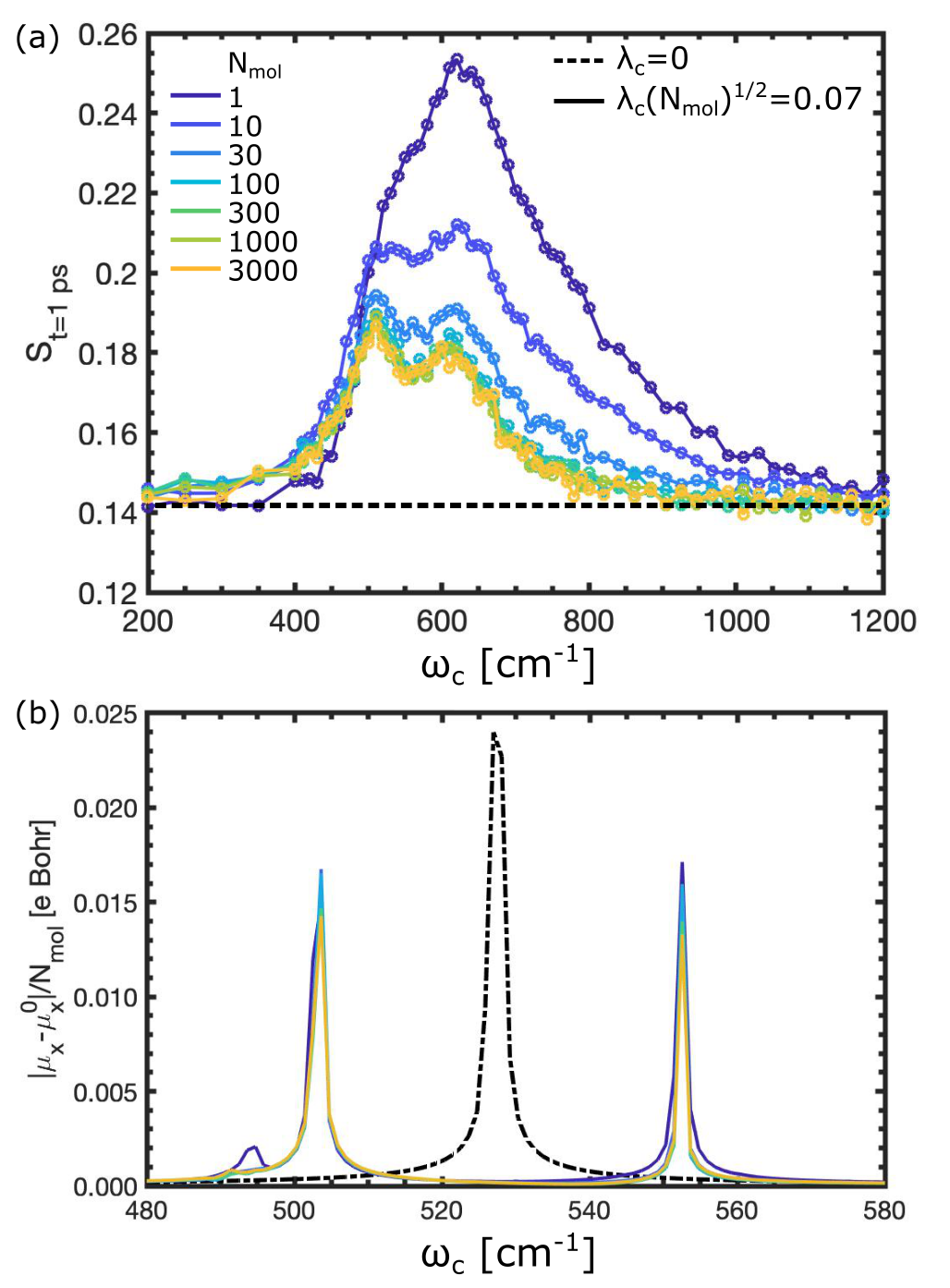}
\caption{\textbf{(a)} Cavity frequency-dependent survival probability after 1 ps for constant $\lambda_\mathrm{c} (N_\mathrm{mol})^{1/2}=0.07$ a.u. with aligned molecules. The curves converge for $N_\mathrm{mol}> 10^2$, suggesting that in the limit of many molecules, $\lambda_\mathrm{c} (N_\mathrm{mol})^{1/2}$ dictates the effect of collective vibrational coupling on chemical reactivity for aligned molecules. \textbf{(b)} Dipolar spectra for this limit normalized to the number of molecule $N_\text{mol}$. The collective Rabi splitting in the dipolar moment spectra scales proportionally with $\lambda_\mathrm{c}(N_\mathrm{mol})^{1/2}$. Here, $\lambda_\mathrm{c}(N_\mathrm{mol})^{1/2}=0.07$ a.u., resulting in a collective Rabi splitting of $\sim50$ cm$^{-1}$.
}
\label{fig:conserveLambdacSqrtNmol}
\end{figure}

Next, we empirically discover an asymptotic scaling law for the onset of reaction rate modification by collective vibrational strong coupling in Fig. \ref{fig:conserveLambdacSqrtNmol}(a). Here, we show the cavity frequency $\omega_\mathrm{c}$-dependent survival probability after 1 ps for varying $N_\mathrm{mol}$ from 1 to 3000 while holding $\lambda_\mathrm{c}(N_\mathrm{ mol})^{1/2}$ constant at 0.07 a.u. Notably, the curves converge as $N_\mathrm{mol}> 10^2$, suggesting that in the limit of many molecules, $\lambda_\mathrm{c}(N_\mathrm{mol})^{1/2}$ dictates the effect of collective vibrational coupling on chemical reactivity for aligned molecules. 
For the case of $N_\mathrm{mol}=1$, we find a broad resonance between 500-700 cm$^{-1}$. Increasing the number of molecules, we find a double peak structure at 520 cm$^{-1}$ and 600 cm$^{-1}$ in the survival probability $S$ reminiscent of the double peaks in the vibrational infrared spectrum for long-lived trajectories of this molecular model, as shown in Ref.~\citenum{Wang2021CavityIVR} in the limit of a single molecule. 
As we demonstrate in the Appendix, it appears that smaller $N_\mathrm{mol}\sim 1-10^2$ have a larger effective molecule-cavity couplings than for $N_\mathrm{mol}> 10^2$ for the same $\lambda_\mathrm{c}(N_\mathrm{mol})^{1/2}$ leading to the broad resonance in the survival probability. This is due to the sampling, where each initial state for a molecule in an ensemble is randomly sampled separately. When the initial states are identical within an ensemble, we find that the curves overlay perfectly even for smaller $N_\mathrm{mol}\sim 1-10^2$. In Appendix \ref{app:scalingrelation}, we also demonstrate how to precisely extract this scaling law with a range of $\lambda_\mathrm{c}(N_\mathrm{mol})^{1/2}$ values.

Notably, the scaling of the survival probability with $\lambda_\mathrm{c}(N_\mathrm{mol})^{1/2}$ is identical to the scaling of the collective Rabi splitting with the number of molecules, or the change in frequency of the vibrational dynamics of the collectively coupled molecules inside vs. outside a cavity due to interactions between the molecular vibrations and the cavity. We numerically demonstrate this scaling in Fig. \ref{fig:conserveLambdacSqrtNmol}(b), where we plot the $N_\mathrm{mol}$-normalized dipolar moment spectrum $|\mu_x-\mu_x^0|(\omega)/N_\mathrm{mol}=|\int \mathrm{d}t (\mu_x(t)-\mu_x^0)\mathrm{exp}(-2\pi\mathrm{i}\omega t)|/N_\mathrm{mol}$. This expression corresponds to the one-sided Fourier transform of the permanent dipole moment $\mu_x$ as the displacements $q_i$ and momenta $p_i$ propagate under $H$.

In the present system, $\lambda_\mathrm{c}(N_\mathrm{mol})^{1/2}=0.07$ a.u. results in a collective Rabi splitting of $\sim50$ cm$^{-1}$, or about 10\% of the vibrational mode frequency. This regime is regularly achieved in experimental setups of vibrational polariton chemistry \cite{Thomas2016, Thomas2019} and, due to cavity and molecular losses, corresponds approximately to the onset of the strong coupling regime from the weak coupling regime. 

\begin{figure}[!tbhp]
\centering
\includegraphics[width=1.0\linewidth]{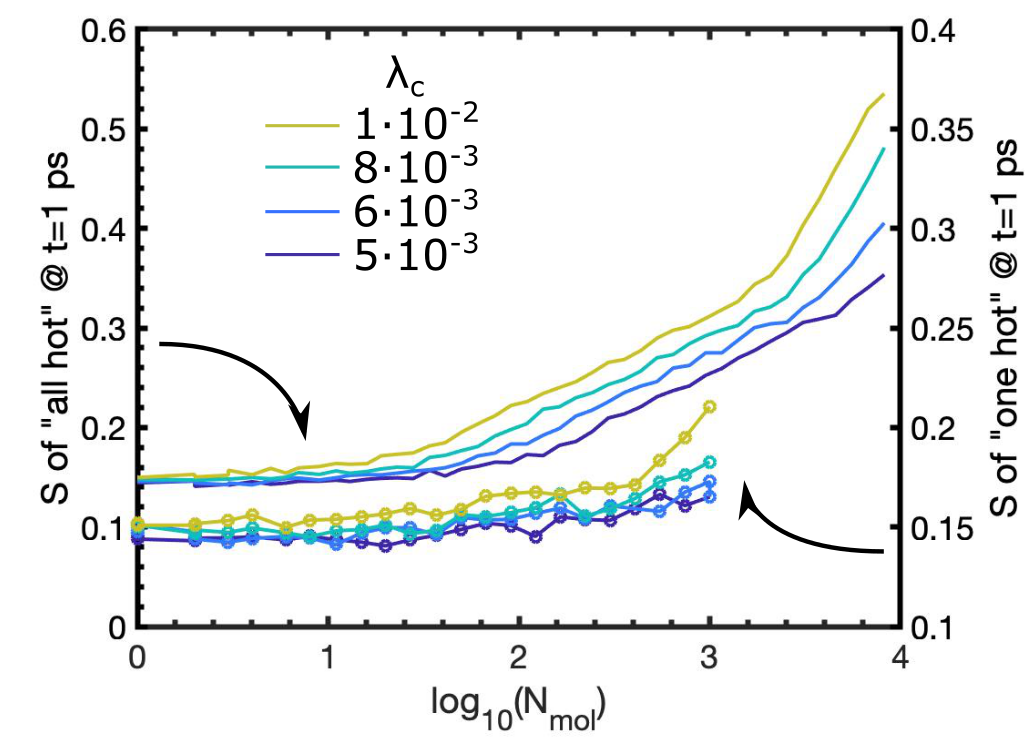}
\caption{Cavity-modified survival probability $S$ after $t=1$ ps for conditions closer to thermal reactions with varying $\lambda_\mathrm{c}$ and constant $\omega_\mathrm{c}=520$ cm$^{-1}$. We compare two cases: `one hot' (dotted with $y$-axis on the right) where each ensemble has one hot molecule among $N_\mathrm{mol}-1$ cold molecules, and `all hot' (lines with $y$-axis on the left) where each molecule in an ensemble is hot. The arrows point to the corresponding $y$-axis. Although the relative change in reaction rate with increasing $N_\mathrm{mol}$ is smaller for the `one hot' case than for the `all hot' case, we observe similar changes in reaction rate with increasing $N_\mathrm{mol}$.
}
\label{fig:oneHot}
\end{figure}

In Fig. \ref{fig:Nmol} and Fig. \ref{fig:conserveLambdacSqrtNmol}, we assume the molecules, each with energy of 34 kcal/mol $>$ the dissociation energy $D$, are aligned along the field of the cavity mode with frequency $\omega_\mathrm{c}$. In thermal reactions of interest, however, at any given moment, only a few molecules are `hot' enough to react among a bath of colder molecules near room temperature. Therefore, in Fig. \ref{fig:oneHot}, we explore whether the empirically observed scaling law is robust to this change. Specifically, we compare the survival probability after 1 ps for a single hot molecule (34 kcal/mol $>$ dissociation energy $D$) among $N_\mathrm{mol}-1$ cold (1 kcal/mol$\sim$43 meV, or on the order of room temperature) molecules vs. ensembles where every molecule is initialized with 34 kcal/mol, or `one hot' vs. `all hot,' respectively. The former more closely corresponds a thermal reaction at low temperatures, where there are comparatively few hot molecules capable of dissociating among the many cold molecules. Notably, the `one hot' case still results in $N_\mathrm{mol}$-dependence, although the relative change in reaction rate is smaller than for the `all hot' case. This difference is likely due to the smaller vibrational mode amplitudes of cold molecules compared to that of hot molecules, leading to relatively smaller changes in dipole moment and less coupling with the cavity. Nevertheless, this result suggests that the observed scaling law between $\lambda_\mathrm{c}$ and $N_\mathrm{mol}$ still holds for hot reacting molecules in a bath of cold molecules.

\begin{figure}[!tbhp]
\centering
\includegraphics[width=1.0\linewidth]{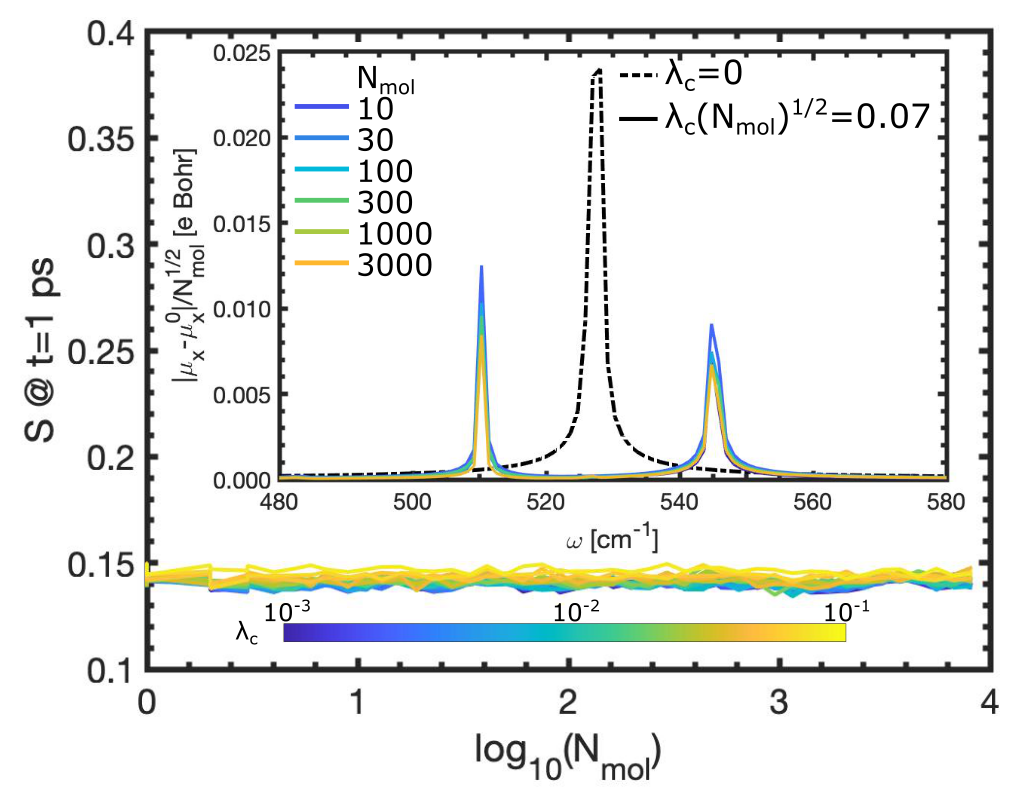}
\caption{Cavity-modified chemical reactivity of randomly oriented molecules, specifically $N_\mathrm{mol}$-dependent survival $S$ after $t=1$ ps with varying cavity strength $\lambda_\mathrm{c}$ and $\omega_\mathrm{c}=520$ cm$^{-1}$. Notably, there is no clear $\lambda_\mathrm{c}$-dependent onset of $N_\mathrm{mol}$ dependence. Inset: Dipolar spectra for this limit normalized to the square-root of the number of molecule ${N}_\mathrm{mol}^{1/2}$. Here, $\lambda_\mathrm{c}(N_\mathrm{mol})^{1/2}=0.07$ a.u., resulting in a collective Rabi splitting of $\sim35$ cm$^{-1}$.}
\label{fig:rotation}
\end{figure}

In addition, in most experimental studies of vibrational polariton chemistry, the reactants are in the liquid phase with weak orientational preferences \cite{Li2020b}. Therefore, we randomly orient the molecules by uniformly sampling $\theta_n$ from 0 to 2$\pi$, randomly initialize each molecule with an energy of 34 kcal/mol, and study the $N_\mathrm{mol}$-dependent survival probability $S$ for varying $\lambda_\mathrm{c}$ in Fig. \ref{fig:rotation}(a). In this setup, we observe no visible changes in the chemical reaction rate. We rationalize this qualitative difference in scaling between aligned and random orientations by comparing the $N_\mathrm{mol}$-dependent dipolar spectrum of aligned molecules in Fig. \ref{fig:conserveLambdacSqrtNmol}(b) and randomly oriented molecules in Fig. \ref{fig:rotation}(b). First, we note that the effective $\lambda_\mathrm{c}$ in the randomly oriented case is scaled down by a factor of $\sqrt{2}$ relative to the aligned case, as can be seen in the Rabi splitting of 35 cm$^{-1}$ vs. 50 cm$^{-1}$ in the aligned case; this scaling can be straightforwardly derived from, for instance, the Dicke model of randomly oriented dipoles \cite{Keeling2018}. Second and also important for observing collective enhancements of chemical reactivity, while the magnitude of the total dipolar spectrum for the aligned molecules scales proportionally with increasing $N_\mathrm{mol}$, the magnitude of the total dipolar spectrum for the randomly oriented molecules scales only with $N_\mathrm{mol}^{1/2}$. Recalling that the cavity mode $Q_\mathrm{c}$ couples to the total dipole moment, we find that this weaker scaling under random orientation and random initialization that results in weaker per-molecule light-matter coupling compared to the aligned setup is evidently not capable of modifying chemical reactivity toward large $N_\mathrm{mol}$. Similar reduced scaling for random sampling of initial states has been found for intermolecular energy transfer in Ref. \citenum{Li2021e}. We note that we can attribute the reduced scaling to the random sampling of initial states, as we show in the Appendix in Fig.~\ref{fig:randomlyOrientedCoherentlyKicked}, where coherent initial states of randomly oriented molecules also show scaling with $N_\mathrm{mol}$ as in the aligned case.

\section{Conclusions}
In summary, we study chemical reactivity under collective vibrational strong coupling, specifically unimolecular dissociation of a classical and anharmonic triatomic model. We find that with increasing $N_\mathrm{mol}$ in the ensemble, the reaction rate decreases due to collective enhancement of vibrational energy transfer to the cavity as the molecules drive the cavity displacement. We empirically find a scaling relation for the onset of $N_\mathrm{mol}$-dependence of this effect. Extrapolating this analysis to experimental conditions of smaller $\lambda_\mathrm{c}$ and larger $N_\mathrm{mol}$, we believe that such effects may be observed in experimentally realizable infrared cavities for aligned reactants, although further study is needed to explore beyond the limit of low cavity loss and intermolecular interactions. However, because changes to the chemical reactivity are not observed for randomly oriented molecules under the significantly smaller cavity strengths of experimentally realizable cavities, we remain wary of the applicability of this theory to previous experiments of vibrational polariton chemistry with liquid phase reactants. Future studies should explore whether molecules can be partially aligned in realistic, micron-scale cavities and if alignment is necessary to realize collectively-modified chemical reactivity.

\section*{Acknowledgements}
D.S.W. and J.F. contributed equally to this manuscript.
The authors acknowledge valuable discussions with Arkajit Mandal and Tom\'a\v{s} Neuman. D.S.W. is an NSF Graduate Research Fellow. S.F.Y. would like to acknowledge funding by the DOE, AFOSR, and NSF for supporting research into molecular ensembles, development of numerical methods, and development of theoretical formalism, respectively. Calculations were performed using the computational facilities of the Flatiron Institute. The Flatiron Institute is a division of the Simons Foundation. 

\appendix

\section{Dipole moment} \label{app:dipolemoment}

We parametrize the permanent dipole moment as 
\begin{subequations}
    \begin{align}
        \bm{\mu}_n & =[\cos{(\theta_n)} \mu_{\alpha,n}+\sin{(\theta_n)} \mu_{\beta,n}] \bm{\hat{x}} \nonumber \\ & + [\sin{(\theta_n)} \mu_{\alpha,n}-\cos{(\theta_n)} \mu_{\beta,n}] \bm{\hat{y}} \\
        & = \mu_x \bm{\hat{x}} + \mu_y \bm{\hat{y}},
    \end{align}
\end{subequations}
where $\mu_{\alpha,n}=A\cos{(q_{n,3}/2)}[\mathcal{F}(q_{n,1}) + \mathcal{F}(q_{n,2})]$, $\mu_{\beta,n}=A\sin{(q_{n,3}/2)}[\mathcal{F}(q_{n,1}) - \mathcal{F}(q_{n,2})]$, $\mathcal{F}(q_{n,i})=q_{n,i}e^{-(q_{n,i} - q_i^0)^2/(2\sigma^2)}$, $\sigma=1$ Bohr controls the envelope width, $\theta_n$ is the angle between the molecular axis and $x$-axis, and $A=0.4$ e$\cdot$Bohr is such that the equilibrium permanent dipole moment $\bm{\mu}_0$ in the $x$-direction is 1 e$\cdot$Bohr. Importantly, the dipole moment approaches zero with increasing $q_1$ and $q_2$, physically analogous to dissociation into neutral species, to obviate challenges with the dipole approximation after dissociation.

\section{Initial states} \label{app:initialstates}
We briefly review the method of generating initial states, similarly to previous work in  Refs.~\citenum{Karmakar2020, Wang2021CavityIVR} for a single molecule.

For a given molecule $n$ with energy $E_\mathrm{mol}$, we compute the minimum $Q_{n,\mathrm{min}}$ and maximum $Q_{n,\mathrm{max}}$ possible values of $Q_n\in \{q_{n,1}, q_{n,2}, p_{n,1}, p_{n,2}, p_{n,3}\}$ corresponding to the cases when $E_\mathrm{mol}$ is contained entirely within the potential or kinetic energy associated with $Q_n$. Then, we generate all possible initial states from uniformly spaced arrays with $N$ values between $Q_{n,\mathrm{min}}$ and $Q_{n,\mathrm{max}}$. For each of these states with energy less than $E_\mathrm{mol}$, we choose $q_{n,3}-q_{n,3}^0$ that results in a total energy of $E_\mathrm{mol}$. To obtain converged results in this study, we choose $N$ between 15 and 23, where larger values are necessary for survival probabilities and smaller values are sufficient for Fourier spectra and time-dependent energy curves. For each ensemble, we then randomly select from these states for each of $N_\mathrm{mol}$ initial states in an ensemble. We propagate $N_\mathrm{mol} \cdot N_\mathrm{traj} > 10^4$ total trajectories for each curve in Fig. \ref{fig:Nmol}(a) and each data point in Figs. \ref{fig:conserveLambdacSqrtNmol}, \ref{fig:oneHot}, \ref{fig:rotation}, \ref{fig:resonance}, and \ref{fig:scaling}(a). The cavity is initialized to set the contribution from the field Hamiltonian $H_\mathrm{F}$ to 0. The effects of different initial conditions for the cavity mode are studied in further detail in Ref.~\citenum{Wang2021CavityIVR}. For the case of randomly oriented molecules in Fig. \ref{fig:rotation}, for each ensemble of $N_\mathrm{mol}$ molecules, we fix the randomly selected orientations $\theta_n$ and sample randomly selected initial states over multiple trajectories.

\section{Sampling schemes} \label{app:sampling}
\begin{figure}[!tbhp]
\centering
\includegraphics[width=\linewidth]{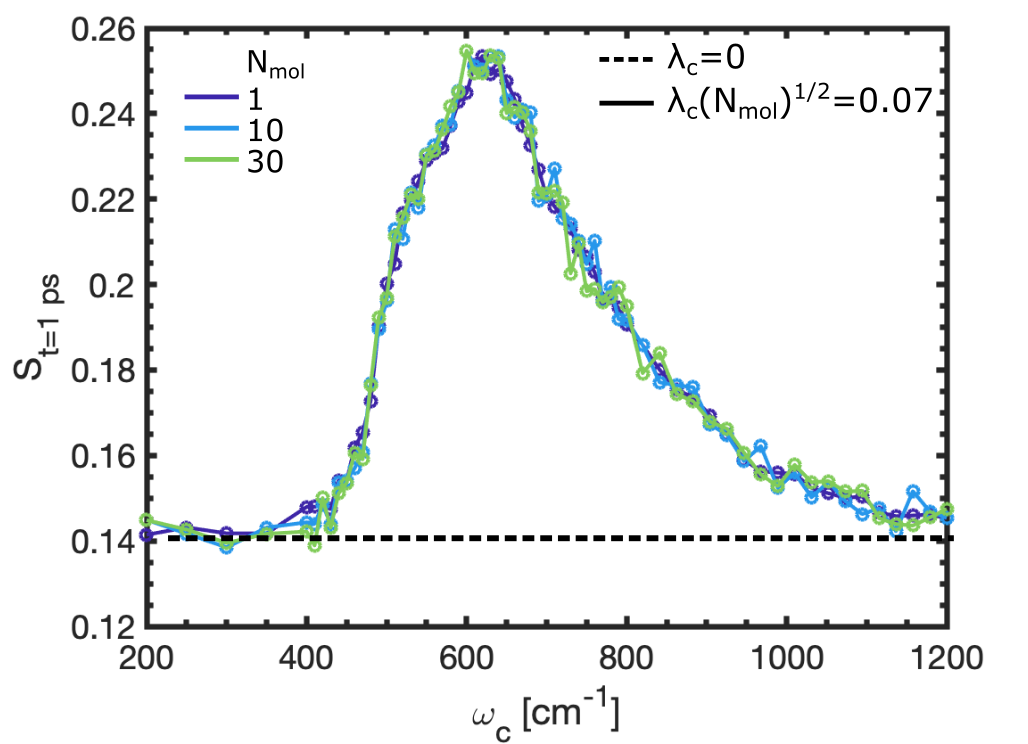}
\caption{Cavity frequency-dependent chemical reactivity for constant $\lambda_\mathrm{c} (N_\mathrm{mol})^{1/2}=0.07$ a.u. with aligned molecules with coherent initialization, \textit{i.e.} each molecule in an ensemble has the same initial state. The curves are identical for increasing $N_\mathrm{mol}$ from $N_\mathrm{mol}=1$, suggesting that under these initialization conditions, $\lambda_\mathrm{c} (N_\mathrm{mol})^{1/2}$ entirely dictates the effect of collective vibrational coupling on chemical reactivity for aligned molecules.}
\label{fig:conserveLambdacSqrtNmolCoherent}
\end{figure}

Here we explore the effects of different initialization procedures. First, we study coherent initialization for aligned molecules, motivated by the observation that in Fig. \ref{fig:conserveLambdacSqrtNmol}, the effective coupling for low $N_\mathrm{mol} \sim 1-10^2$, despite constant $\lambda_\mathrm{c}(N_\mathrm{mol})^{1/2}$ for all $N_\mathrm{mol}$, appears to be higher than for large $N_\mathrm{mol}> 10^2$. We show that this difference can be attributed to the initialization conditions, where in Fig. \ref{fig:conserveLambdacSqrtNmol}, each molecule in an ensemble of $N_\mathrm{mol}$ molecules is initialized with a different random state, whereas in Fig. \ref{fig:conserveLambdacSqrtNmolCoherent}, each molecule in an ensemble is initialized with the same state. In this latter case, we see that the cavity-modified reactivity is entirely determined by $\lambda_\mathrm{c}(N_\mathrm{mol})^{1/2}$.

\begin{figure}[!tbhp]
\centering
\includegraphics[width=\linewidth]{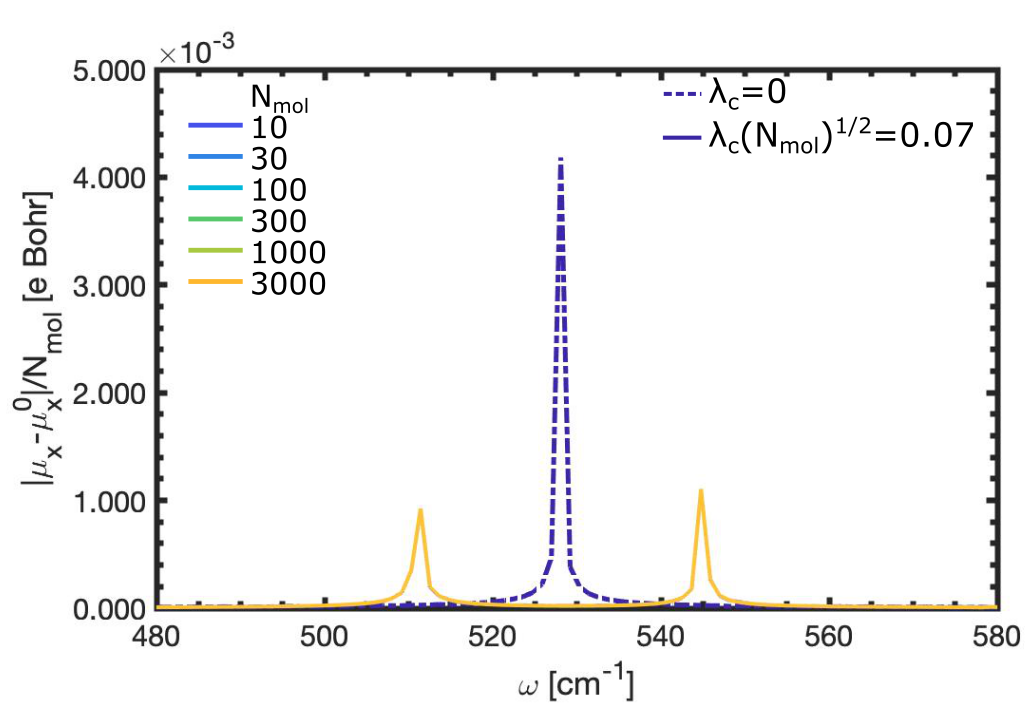}
\caption{$N_\mathrm{mol}$-normalized total dipolar spectrum for randomly oriented and coherently kicked molecules in a cavity. The magnitude of the spectrum scales proportionally with $N_\mathrm{mol}$, just as in the aligned case in Fig. \ref{fig:conserveLambdacSqrtNmol} that exhibits collectively-enhanced chemical reactivity. All curves for $\lambda_\mathrm{c}(N_\mathrm{mol})^{1/2}=0.07$ a.u. lie on top of each other.}
\label{fig:randomlyOrientedCoherentlyKicked}
\end{figure}

We also explore how the scaling of the total dipolar spectrum for randomly oriented molecules that are initialized, not randomly as in the main text, but coherently ``kicked" as in a linear response calculation. Specifically, we displace $q_{1,n}$ and $q_{2,n}$ from the vibrational ground state by $\epsilon \mathrm{cos}(\theta_n)$ for all $n$ where $\epsilon$ is a small value. In Fig. \ref{fig:randomlyOrientedCoherentlyKicked} we find that the total dipolar spectrum scales with $N_\mathrm{mol}$, as in the aligned case that exhibits collectively-enhanced chemical reactivity. 

\section{Molecule number-dependent cavity resonance}

\begin{figure}[!tbhp]
\centering
\includegraphics[width=1.0\linewidth]{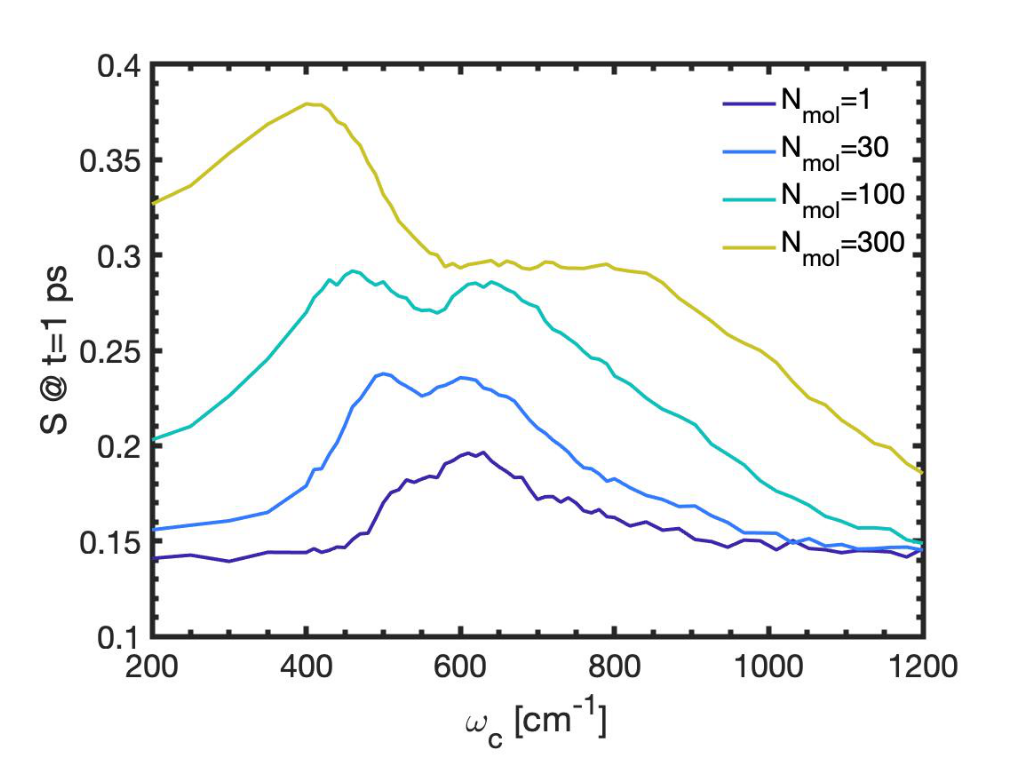}
\caption{Cavity frequency $\omega_\mathrm{c}$-dependent survival probability $S$ after 1 ps for varying $N_\mathrm{mol}$ and $\lambda_\mathrm{c}=10^{-2}$ a.u.. As expected, we observe a resonant effect for $N_\mathrm{mol}=1$ at $\omega_\mathrm{c}=520$ cm$^{-1}$ and a weaker one at 600 cm$^{-1}$. As $N_\mathrm{mol}$ increases, however, the resonant cavity frequencies shift.
}
\label{fig:resonance}
\end{figure}

In experiments, the cavity frequency can be varied, leading to the discovery of the resonant effect where the change in reaction rate is largest at certain frequencies observed to coincide with particular vibrational modes \cite{Thomas2016, Thomas2019, Pang2020, Sau2021} and the original motivation for studying cavity-mediated intramolecular vibrational energy redistribution \cite{Schafer2021,  Wang2021CavityIVR}. Therefore, in Fig. \ref{fig:resonance}, we explore how the resonance frequencies change with increasing molecule-cavity coupling. In particular, we compute the survival probability $S$ after 1 ps with $\lambda_\mathrm{c}=10^{-2}$ a.u. for $N_\mathrm{mol}=\{1, 30, 100, 300\}$, where the chemical reaction rate is slowed with increasing $\lambda_\mathrm{c}$ due to cavity-mediated vibrational energy transfer. As expected from Ref.~\citenum{Wang2021CavityIVR}, we observe resonances for $N_\mathrm{mol}=1$ at $\omega_\mathrm{c}=520$ cm$^{-1}$ and a weaker one at 600 cm$^{-1}$. Interestingly, as $N_\mathrm{mol}$ increases, the resonant cavity frequencies shift with increasing difference between the two frequencies.

\begin{figure}[!tbhp]
\centering
\includegraphics[width=\linewidth]{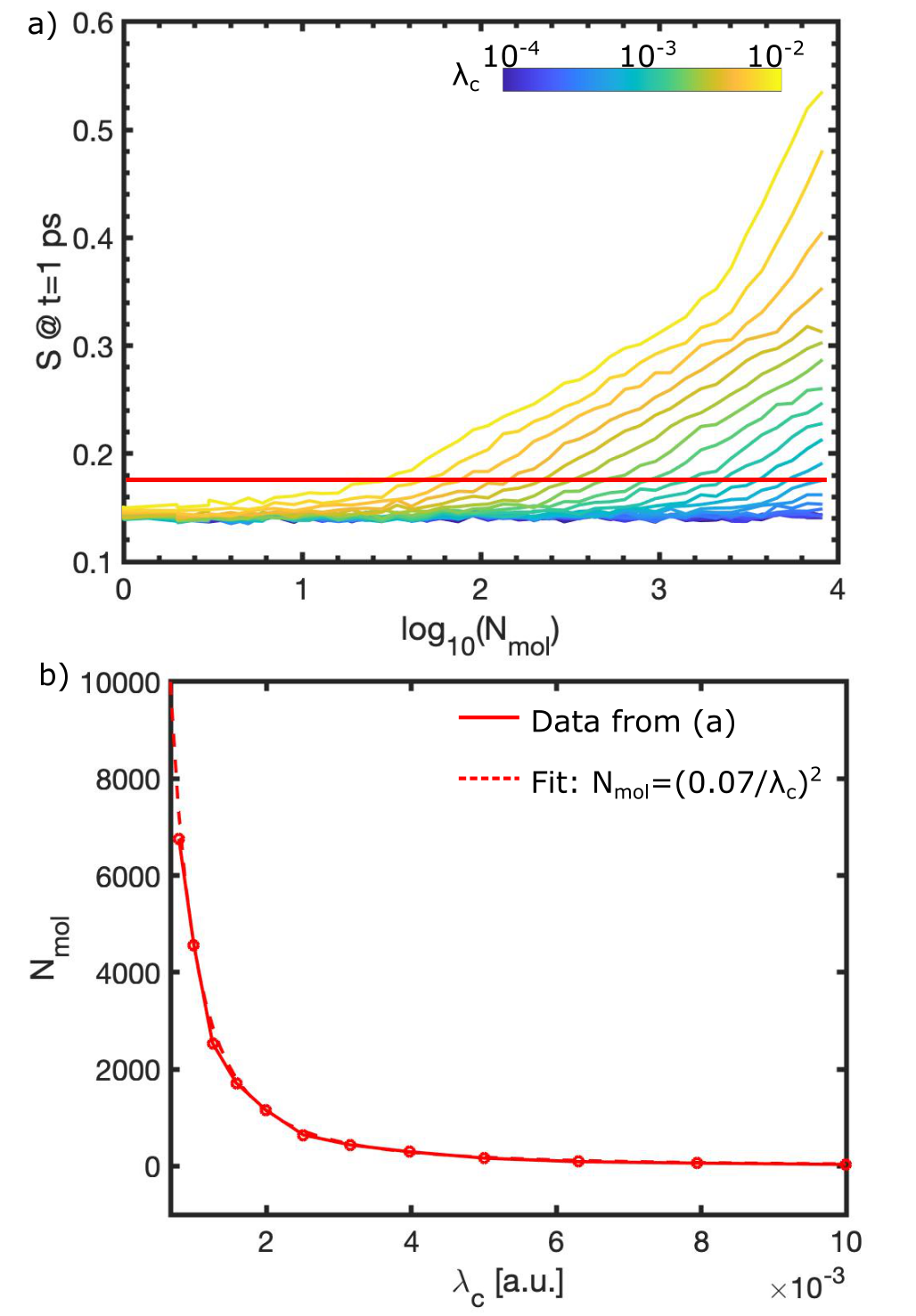}
\caption{\textbf{(a)} Survival probability after 1 ps for constant $\lambda_\mathrm{c}$ and increasing $N_\mathrm{mol}$. \textbf{(b)} Minimum $N_\mathrm{mol}$, above which the reaction rate exhibits $N_\mathrm{mol}$-dependence, for varying $\lambda_\mathrm{c}$, which correspond to the points intersected by the red line in (a). We overlay a fitting curve $N_\mathrm{mol}=(0.07/ \lambda_\mathrm{c})^2$ that agrees well with the numerical data.
}
\label{fig:scaling}
\end{figure}

\section{Scaling relation for large $N_\mathrm{mol}$} \label{app:scalingrelation}

Here, we demonstrate how we extract the scaling law $\lambda_\mathrm{c}(N_\mathrm{mol})^{1/2}=0.07$ a.u.. In Fig. \ref{fig:scaling}(a) we plot the survival probability $S$ after 1 ps of propagation, at which time the rate slowdown due to cavity-mediated intermolecular vibrational energy transfer can be clearly observed, as a function of $N_\mathrm{mol}$ for varying $\lambda_\mathrm{c}$. As $\lambda_\mathrm{c}$ decreases, the onset of rate slowdown occurs at increasing $N_\mathrm{mol}$. Then, in Fig. \ref{fig:scaling}(b), we plot the threshold $N_\mathrm{mol}$ at which the reaction slows down with increasing $N_\mathrm{mol}$ as a function of $\lambda_\mathrm{c}$ and overlay a fitting curve of $\lambda_\mathrm{c}(N_\mathrm{mol})^{1/2}=7 \cdot 10^{-2}$ a.u.

\FloatBarrier

\typeout{}

\end{document}